\documentclass[aps,twocolumn,showpacs,citeautoscript,reprint,prb,floatfix]{revtex4-1}

\usepackage{bm}
\usepackage{graphicx}
\usepackage{amsmath}
\usepackage{amsfonts}
\usepackage{bbold}
\usepackage{amssymb}
\usepackage{siunitx}
\usepackage{color}

\newcommand{\sgn}{\mathrm{sgn\,}}

\begin{document}

\title{Topologically protected states in $\delta$--doped junctions with band inversion}

\author{A. D\'{i}az-Fern\'{a}ndez}

\affiliation{GISC, Departamento de F\'{\i}sica de Materiales, Universidad Complutense, E--28040 Madrid, Spain}

\author{N. del Valle}

\affiliation{GISC, Departamento de F\'{\i}sica de Materiales, Universidad Complutense, E--28040 Madrid, Spain}

\author{E. D\'{\i}az}

\affiliation{GISC, Departamento de F\'{\i}sica de Materiales, Universidad Complutense, E--28040 Madrid, Spain}

\author{F. Dom\'{i}nguez-Adame}

\affiliation{GISC, Departamento de F\'{\i}sica de Materiales, Universidad Complutense, E--28040 Madrid, Spain}

\begin{abstract}

A topological boundary can be formed at the interface between a trivial and a topological insulator. The difference in the topological index across the junction leads to robust gapless surface states. Optical studies of these states are scarce in the literature, the reason being the difficulty to isolate their response from that of the bulk. In this work, we propose to deposit a $\delta$ layer of donor impurities in close proximity to a topological boundary to help detecting gapless surface states. As we will show, gapless surface states are robust against this perturbation and they enhance intraband optical transitions as measured by the oscillator strength. These results allow to understand the interplay of surface and bulk states in topological insulators.


\end{abstract}


\maketitle

\section{Introduction}

Topologically-protected surface states naturally arise at the boundary between a topological and a trivial insulator or vacuum~\cite{Hasan10,Bansil16,Tchoumakov17}. The robustness of these states stems from discrete symmetries of the bulk. As a result, topological insulators are often included in the category of symmetry-protected topological phases, as opposed to topologically ordered phases, like the fractional quantum Hall states. This classification can be understood in terms of short- and long-range entanglement of the ground state, respectively~\cite{Wen17,Senthil15}. Among the vast myriad of symmetry-protected topological phases that are known to date, topological crystalline insulators~\cite{Ando15} and three-dimensional topological insulators~\cite{Hasan10} are particularly relevant. The former are protected by crystalline symmetries, such as mirror symmetry, and can be characterized by a topological invariant, namely, a mirror Chern number~\cite{Hsieh12}. Specific examples with experimental support of these topological crystalline insulators are Pb$_{1-x}$Sn$_{x}$Te~\cite{Hsieh12,Assaf16,Hasan12} and Pb$_{1-x}$Sn$_{x}$Se~\cite{Dziawa12}. These materials shift from being trivial insulators to topological crystalline insulators as the Sn fraction, $x$, is increased. The evolution from trivial to topological corresponds to a band closure in the bulk at the $L$ points of the Brillouin zone when a critical value of $x$ is reached. The bands that undergo band inversion are the $L_6^{+}$ and $L_6^{-}$. Upon increasing $x$ further, the gap reopens. This is a signature of a topological phase transition.

On the other hand, the aforementioned three-di\-men\-sion\-al topological insulators are protected by somehow more subtle symmetries. The first experimental discovery was Bi$_{1-x}$Sb$_x$ in 2008~\cite{Hsieh08}. However, this material proved to have a rather complicated surface structure and a comparably small energy gap. A year later, a family of so-called \emph{second generation materials}~\cite{Moore09} was discovered, among which Bi$_2$Se$_3$ stands out due to its remarkable properties, such as the possibility to exploit its topological nature at room temperature~\cite{Hasan10}. Time reversal and parity inversion symmetries are responsible for its topological protection. A two-band approximation reminiscent of the times of Volkov and Pankratov~\cite{Volkov85,Korenman87,Agassi88,Pankratov90,Adame94,Kolesnikov97} can be put forward to describe these two kinds of topological insulators~\cite{Hsieh12,Kane12,Tchoumakov17}. A $\mathbb{Z}_2$ topological index can be defined by the sign of the Dirac mass~\cite{Kane12}, which in this case corresponds to half the energy gap. A topological boundary that hosts surface states can be grown by having opposite invariants on each side of the boundary. The resulting surface states are Dirac cones living within the fundamental gap. Remarkably, the Fermi velocity of these cones can be dynamically tunned by external fields~\cite{Diaz-Fernandez17a,Diaz-Fernandez17b,Diaz-Fernandez17c,Diaz-Fernandez18}.


The existence of topological surface states has been probed by angle-resolved photoemission spectroscopy~\cite{Bianchi10,Bianchi11,Chen12,Xu12}, scanning tunneling microscopy~\cite{Mann13}, electron transport~\cite{Inhofer17} and Shubnikov-de Haas oscillations~\cite{Veyrat15} (see Ref.~\onlinecite{Ortmann15} for a comprehensive review). In contrast, optical studies are scarce in the literature~\cite{Rahim17,MoscaConte2017} since it is not straightforward to isolate the optical response of topological surface states from that of the bulk states. In this work we show that this is not necessarily the case. If the population of these surface states is increased, one can expect an enhancement of their optical response. Therefore, in order to better observe the linear optical response of topological surface states, we propose to evaporate during growth a sheet of shallow donor (or acceptor) impurities at a small distance from a band-inverted boundary ($\delta$ doping). We then theoretically study the electronic structure of such a device using a minimal two-band model. Under reasonable assumptions, we obtain a solvable model using the nonlinear Thomas-Fermi (TF) formulation. Subsequently, we show that intraband optical transitions carry information not displayed in a junction between two trivial semiconductors. In the following sections, we will refer to the case of topological crystalline insulators for concreteness, that is, to the aforesaid IV-VI compounds.

\section{Solvable nonlinear Thomas-Fermi formulation}

The system we study in this work is a topological boundary which, as discussed in the introduction, will exhibit topologically-protected surface states within the gap. For our calculations, we shall consider same-sized, aligned gaps. This simplification allows to capture the main physics while keeping the algebra simpler~\cite{Diaz-Fernandez17b}. 

\begin{figure}[ht]
\centering{\includegraphics[width=0.75\columnwidth]{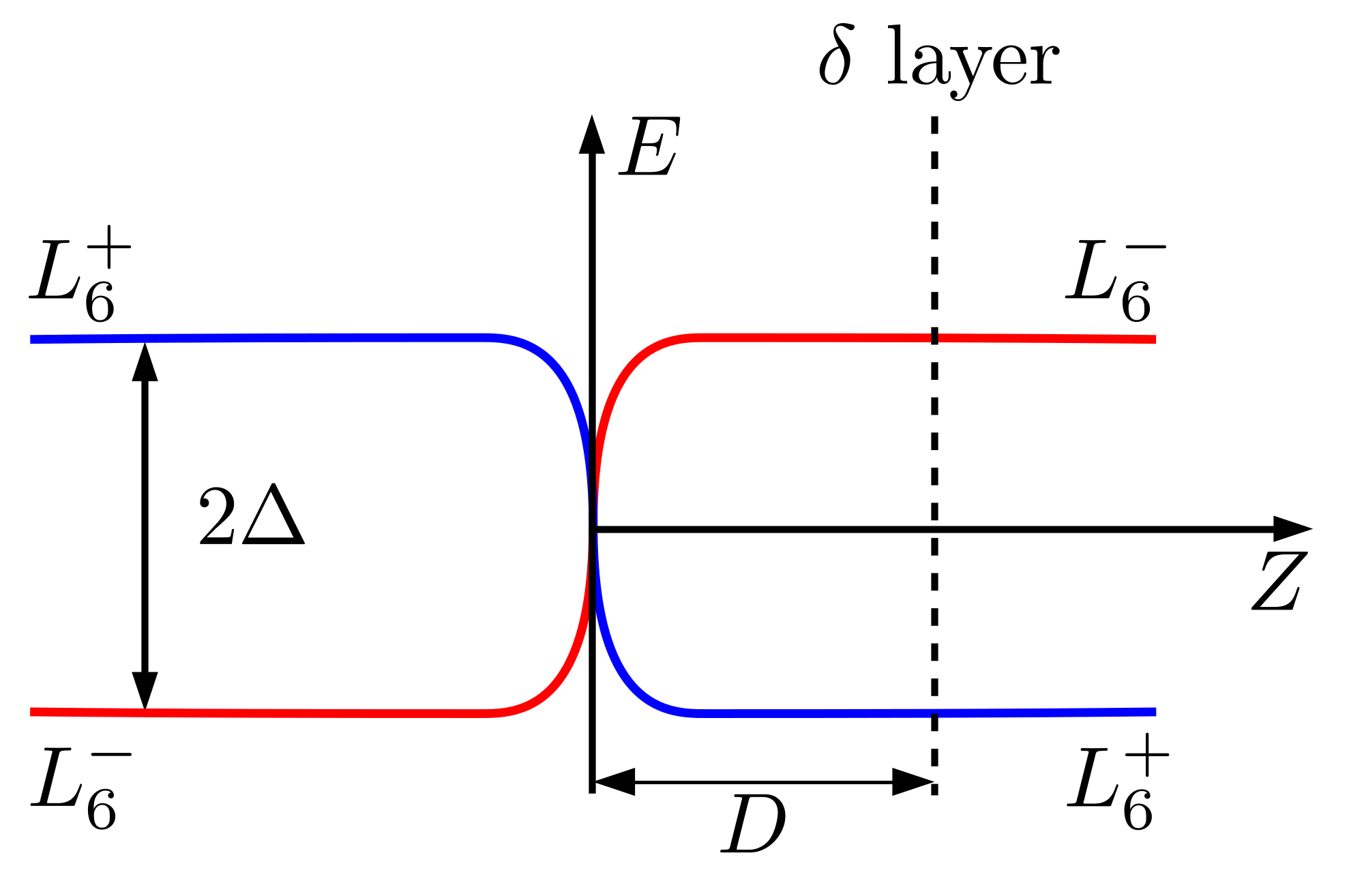}}
\caption{$L_6^{+}$ and $L_6^{-}$ band-edge profile of a band-inverted junction with aligned and same-sized gaps, located at the $XY$ plane. $Z$ indicates the growth direction. The magnitude of the gap is $2\Delta$. A $\delta$ layer of shallow donor impurities is grown at a distance $D$ from the junction.}
\label{fig1}
\end{figure}

To populate these midgap states, we propose to evaporate during growth a $\delta$ layer of shallow donor impurities at a distance $D$ of the junction, as depicted in Fig.~\ref{fig1}. A V-shaped potential is generated at the location of the $\delta$ layer by the ionized donor impurities due to partial screening of the Coulomb potential. Consequently, electron states from the continuum (i.e. the conduction band) are sucked in by this potential and energy quantization results from quantum confinement effects (see Ref.~\onlinecite{Whall92} for a review on $\delta$ doping of semiconductors). We will often refer to this potential as TF well, due to the close analogy to what happens in a square quantum well.

Electrons released from the $\delta$ layer of ionized donor impurities form a two-dimensional electron gas in the vicinity of the layer. Electrons interact with themselves and experience the collective attraction of all ionized impurities. The resulting electronic structure can be calculated in the one-electron approximation, using the local-density functional concept~\cite{Scolfaro94}. The exchange-correlation potential is usually taken in the approximation of Hedin and Lundqvist~\cite{Hedin71} and standard self-consistent numerical methods can be then used~\cite{Koenraad90,Henriques93,Chen93,Cuesta95}. However, the nonlinear TF  formulation of the $\delta$ doping has been proven to be equivalent to the self-consistent (Hartree) model in a wide range of doping densities~\cite{Ioriatti90,Gonzalez94,Kortus94,Mendez94}. The advantage of the TF formulation is that Poisson and Schr\"{o}dinger equations are effectively decoupled and their solution is easier. 

We calculate the space charge potential $V(z)$ ($z$ denotes the spatial coordinate along the growth direction) by means of the TF formulation. The origin of the $z$ coordinate is set at the middle of the $\delta$ layer throughout this section. Neglecting the contribution of a small positive background of ionized acceptors for simplicity, the TF equation reads~\cite{Ioriatti90,Gonzalez94}
\begin{equation}
\frac{d^2V(z)}{dz^2}=-\frac{e^2}{3\pi^2\epsilon}\left[\frac{2m^{*}}{\hbar^2}\Big(E_F-V(z)\Big)\right]^{3/2}
+\frac{e^2}{\epsilon}N_\mathrm{D}(z)\ ,
\label{eq:01}
\end{equation}
where $E_F$ is the Fermi energy, $m^{*}$ is the effective mass and $\epsilon$ is the dielectric constant. When the donor density profile $N_\mathrm{D}(z)$ is assumed to be a $\delta$-function, the nonlinear TF equation can be exactly solved~\cite{Ioriatti90}. Thus, we set $N_\mathrm{D}(z)=n_\mathrm{S}\,\delta(z)$ where $n_\mathrm{S}$ corresponds to the surface density of donors. If the effective Bohr radius $a^{*}=4\pi\epsilon\hbar^2/e^2m^{*}$ and the effective Rydberg energy $\mathrm{Ry}^{*}=\hbar^2/2m^{*}(a^{*})^2$ are taken as the natural units of distance and energy, solution to equation~(\ref{eq:01}) representing neutral structures is given by~\cite{Ioriatti90}
\begin{equation}
V(z)-E_F=-\,\frac{\gamma^2}{(\gamma|z|/a^{*}+\omega)^4}\,\mathrm{Ry}^{*}\ ,
\label{eq:02}
\end{equation}
with $\gamma=2/15\pi$ and $\omega=\left(\gamma^3 /\pi n_\mathrm{S}^{*}\right)^{1/5}$. Here $n_\mathrm{S}^{*}=n_\mathrm{S}(a^{*})^2$ is a dimensionless parameter denoting the number of donors per unit Bohr area. In neutral structures, the above equation implies that $E_F$ lies at the lower edge of the conduction band, far away from the $\delta$ layer.

As it was already noticed by Ioratti~\cite{Ioriatti90}, the Ben Daniel-Duke equation for the envelope function~\cite{Bastard89} with $V(z)$ given by Eq.~(\ref{eq:02}) admits exact analytical solutions in term of Mathieu functions~\cite{Abramowitz72}. However, the determination of the energy levels becomes extremely complex. For this reason, we follow a different route with the aim of seeking a solvable TF model. 

\begin{figure}[ht]
\centering{\includegraphics[width=0.8\columnwidth]{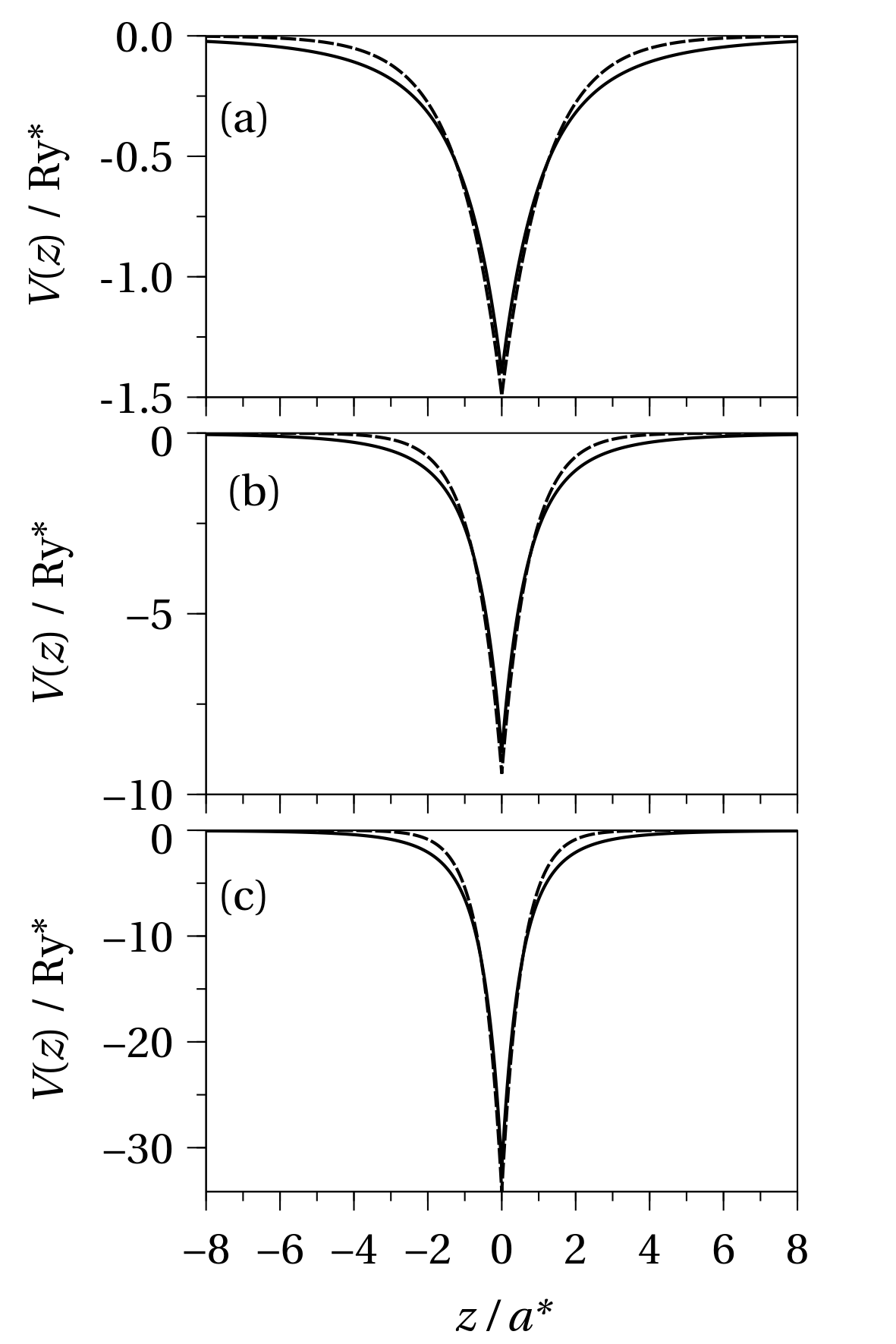}}
\caption{Comparison of the approximate potential $V_\mathrm{app}(z)$ (dashed lines) with the exact TF potential $V(z)$ (solid lines) for different doping levels: (a)~$n_\mathrm{S}^{*}=0.1$, (b)~$n_\mathrm{S}^{*}=1.0$ and (c)~$n_\mathrm{S}^{*}=5.0$.}
\label{fig2}
\end{figure}

The starting point to replace the exact TF potential~(\ref{eq:02}) by an approximate potential $V_\mathrm{app}(z)$ is the charge neutrality condition
\begin{equation}
n_\mathrm{S}=\int_{-\infty}^{\infty}\frac{1}{3\pi^2}\left[\frac{2m^{*}}{\hbar^2}\Big(E_F-V_\mathrm{app}(z)\Big)\right]^{3/2}\mathrm{d}z\ .
\label{eq:03}
\end{equation}
On one side, $V_\mathrm{app}(z)$ should decay fast enough in the limit $|z|\to\infty$ to ensure convergence of the integral. On the other side, close to the origin $V_\mathrm{app}(z)\sim |z|$, similarly to the exact TF potential. These two boundary conditions are met by an approximate potential of the form 
\begin{subequations}
\begin{equation}
V_\mathrm{app}(z)-E_F=-\,v_0\,\exp\left(-\frac{|z|}{\eta a^{*}}\right)\,\mathrm{Ry}^{*}\ ,
\label{eq:04a}
\end{equation}
where the dimensionless parameters $v_0$ and $\eta$ are determined from the charge neutrality condition~(\ref{eq:03})
\begin{align}
\eta&=\left(\frac{3^4\pi}{2^{10}n_\mathrm{S}^{*}}\right)^{1/5}\simeq 
\frac{3}{4}\,\left(n_\mathrm{S}^{*}\right)^{-1/5}\ ,\nonumber \\
v_0&=4\pi \lambda n_\mathrm{S}^{*} \simeq 3\pi \left(n_\mathrm{S}^{*}\right)^{4/5}\ .
\label{eq:04b}
\end{align}
\label{eq:04}
\end{subequations}
Figure~\ref{fig2} shows a comparison of the approximate potential $V_\mathrm{app}(z)$ with the exact TF potential $V(z)$ for different doping levels. We will shortly demonstrate that the approximate potential~(\ref{eq:04}) leads to an exactly solvable two-band model for narrow gap semiconductors~\cite{Adame95,Adame96}. 

\section{Two band model}

A topological boundary can be described by means of the following Dirac-like Hamiltonian~\cite{Agassi88,Pankratov90,Hsieh12,Adame94,Kolesnikov97,Kane12,Goerbig17} 
\begin{equation}
\mathcal{H}=v_F\,{\bm\alpha}\cdot\widehat{\bm{p}}+\frac{1}{2}\,E_{G}(z)\,\beta+V_\mathrm{app}(z)\ ,
\label{eq:05}
\end{equation}
with $D=0$ hereafter (see Fig.~\ref{fig1}). Here ${\bm\alpha}=(\alpha_x,\alpha_y,\alpha_z)$ and $\beta$ denote the usual $4\times 4$ Dirac matrices, $\alpha_{i}=\sigma_x\otimes \sigma_{i}$ and $\beta=\sigma_z\otimes\mathbb{1}_2$, $\sigma_i$ and $\mathbb{1}_n$ being the Pauli matrices and $n\times n$ identity matrix, respectively. Moreover, $v_{F}$ is an interband matrix element having dimensions of velocity and it is assumed scalar, corresponding to isotropic bands around the $L$ point. In order to keep the algebra as simple as possible, we restrict ourselves to the symmetric boundary with same-sized and aligned gaps, as depicted in Fig.~\ref{fig1}. This assumption simplifies the calculations while keeping the underlying physics~\cite{Diaz-Fernandez17b}. Thus, a single and abrupt interface presents the following profile for the magnitude of the gap $E_{G}(z)=2\Delta\,\sgn(z)$, where $\sgn(z)=|z|/z$ is the sign function. Here, the $Z$ axis is parallel to the growth direction $[111]$. 

The Hamiltonian~(\ref{eq:05}) acts upon the envelope function ${\bm F}({\bm r})$, which is a bispinor whose spinor components belong to the $L_{6}^{+}$ and $L_{6}^{-}$ bands. Translational symmetry in the $XY$ plane implies conservation of the in-plane momentum.  Hence, the envelope function can be expressed as ${\bm F}({\bm r})={\bm\chi}(z)\exp(i\,{\bm r}_{\bot}\cdot{\bm p}_{\bot}/\hbar)$, where ${\bm p}_{\bot}$ is the eigenvalue of the in-plane momentum operator $\widehat{\bm{p}}_{\bot}$. It is understood that the subscript $\bot$ in a vector indicates that its $z$--component is zero. It is convenient to introduce the unit of length $d=\hbar v_F/\Delta$ and the following dimensionless magnitudes $\xi=z/d$, $\epsilon=E/\Delta$, $v(\xi)=V_\mathrm{app}(z)/\Delta$ and ${\bm k}={\bm p}_{\bot}d/\hbar$. Since $|\bm{\chi}(z)|^2$ has units of inverse of length, it is also useful to define its dimensionless counterpart as $\bm{\varphi}=\sqrt{d}\,\bm{\chi}$. From the Hamiltonian~(\ref{eq:05}) we get
\begin{subequations}
\begin{equation}
\Big[-i\alpha_z\partial_{\xi}+{\bm\alpha}_{\bot}\cdot {\bm k} +\beta\,\sgn(\xi)+v(\xi)-\epsilon\Big]
{\bm\varphi}(\xi)=0\ ,
\label{eq:06a}
\end{equation}
where now the dimensionless approximate potential can be cast for convenience in the form
\begin{equation}
v(\xi)=-\,\frac{g}{2a}\exp\left(-\,\frac{|\xi|}{a}\right)\ .
\label{eq:06b}
\end{equation}
\label{eq:06}
\end{subequations}
Here $a=\eta a^{*}/d$ and $g=2av_0/\Delta$, where $\lambda$ and $v_0$ are given in Eq.~(\ref{eq:04b}). It is worth mentioning that the same Eq.~(\ref{eq:06}) holds for a $\delta$ doped layer without band-inversion after the substitution $\sgn(\xi) \to 1$. In this case, a closed solution at ${\bm k}=0$ has been reported in Refs.~\onlinecite{Adame95,Adame96}. Following the same procedure described therein, we are able to solve Eq.~(\ref{eq:06a}) in closed form. The transcendent equation for the energy levels in the presence ($\nu=-1$) or absence ($\nu=1$) of band inversion is found to be
\begin{subequations}
\begin{equation}
\big(\lambda \cos\phi-\epsilon\sin\phi\big)^2=\frac{1-\nu}{2}\ ,
\label{eq:07a}
\end{equation}
where $\lambda^2=1+k^2-\epsilon^2$ and $\phi$ is given in terms of Kummer functions~\cite{Abramowitz72} as
\begin{equation}
\phi=g-2\times\mathrm{arg}\Big[M(\lambda a+i\epsilon a,1+2\lambda a,ig)\Big]\ .
\label{eq:07b}
\end{equation}
\label{eq:07}
\end{subequations}
This equation allows us to obtain the dispersion relation $E({\bm k})$ in normal and band-inverted systems. The corresponding envelope functions have to be defined piecewise. We define
\begin{equation}
\delta_{\nu}=\frac{\mu-\mu^{*}}{\mu-\nu\mu^{*}}\ ,
\label{eq:08}
\end{equation}
with $\mu = \left(\epsilon+i\lambda\right)\exp(-i\phi)$ and introduce the following auxiliary functions
\begin{align}
h(\xi) & =  \exp\left(-\lambda\xi-i\,\frac{g}{2}\,e^{-\xi/a}\right)\nonumber  \\
       & \times M\left(\lambda a+i\epsilon a,1+2\lambda a, ige^{-\xi/a}\right)\ , \nonumber  \\
p(\xi) & = e^{i\phi/2}\Theta(\xi)+e^{-i\phi/2}\Theta(-\xi)\ , \nonumber  \\
q(\xi) & = \Theta(\xi)h(\xi)+\Theta(-\xi)h^{*}(-\xi)\ ,
\label{eq:09}
\end{align}
where $\Theta(z)$ is the Heaviside step function. Then, introducing the following two vectors
\begin{align}
\bm{u}(\xi) & = N\,p(\xi)\!\begin{pmatrix} 1 \\[5pt] -\delta_{\nu}ke^{i\theta}\end{pmatrix}\ ,\nonumber \\
\bm{v}(\xi) & = N\,\frac{\epsilon-i\lambda\,\sgn(\xi)}{k^2+1}\,p(\xi)\!\begin{pmatrix} k^2\delta_{\nu}-\rho(\xi) \\[5pt] \big[1+\rho(\xi)\delta_{\nu}\big]\,ke^{i\theta}\end{pmatrix}\ ,
\label{eq:10}
\end{align}
the envelope functions are given by
\begin{equation}
\bm{\varphi}(\xi) = \frac{1}{\sqrt{2}}\begin{pmatrix} \sigma_z & -\sigma_z \\[3pt] \mathbb{1}_2 & \mathbb{1}_2 \end{pmatrix}
\begin{pmatrix}  q(\xi)\,\bm{u}(\xi)\\[3pt] q^{*}(\xi)\,\bm{v}(\xi) \end{pmatrix}\ .
\label{eq:11}
\end{equation}
Here, $k=|\bm{k}|$, $\theta=\arctan(k_y/k_x)$, $\rho(\xi)=1$ if there is no inversion and $\rho(\xi)=\sgn(\xi)$ if there is. $N$ is the normalization constant, which can be obtained from
\begin{equation}
N = \left[4\big(1+k^2|\delta_{\nu}|^2\big)\int_{0}^{\infty}\text{d}\xi~|h(\xi)|^2\right]^{-1/2} \ .
\label{eq:12}
\end{equation}

\section{Results}

We will consider typical values of the parameters in IV-VI compounds throughout this section. Half of the energy gap is about $\Delta=\SI{75}{\milli\electronvolt}$, effective mass $m^{*}=0.05m_0$ ($m_0$ is the free electron mass), relative dielectric constant $\varepsilon_r=15$ and $d=\hbar v_F/\Delta=\SI{4.5}{\nano\meter}$~\cite{Korenman87,Littlewood79}.

\begin{figure}[ht]
\centering{\includegraphics[width=0.8\columnwidth]{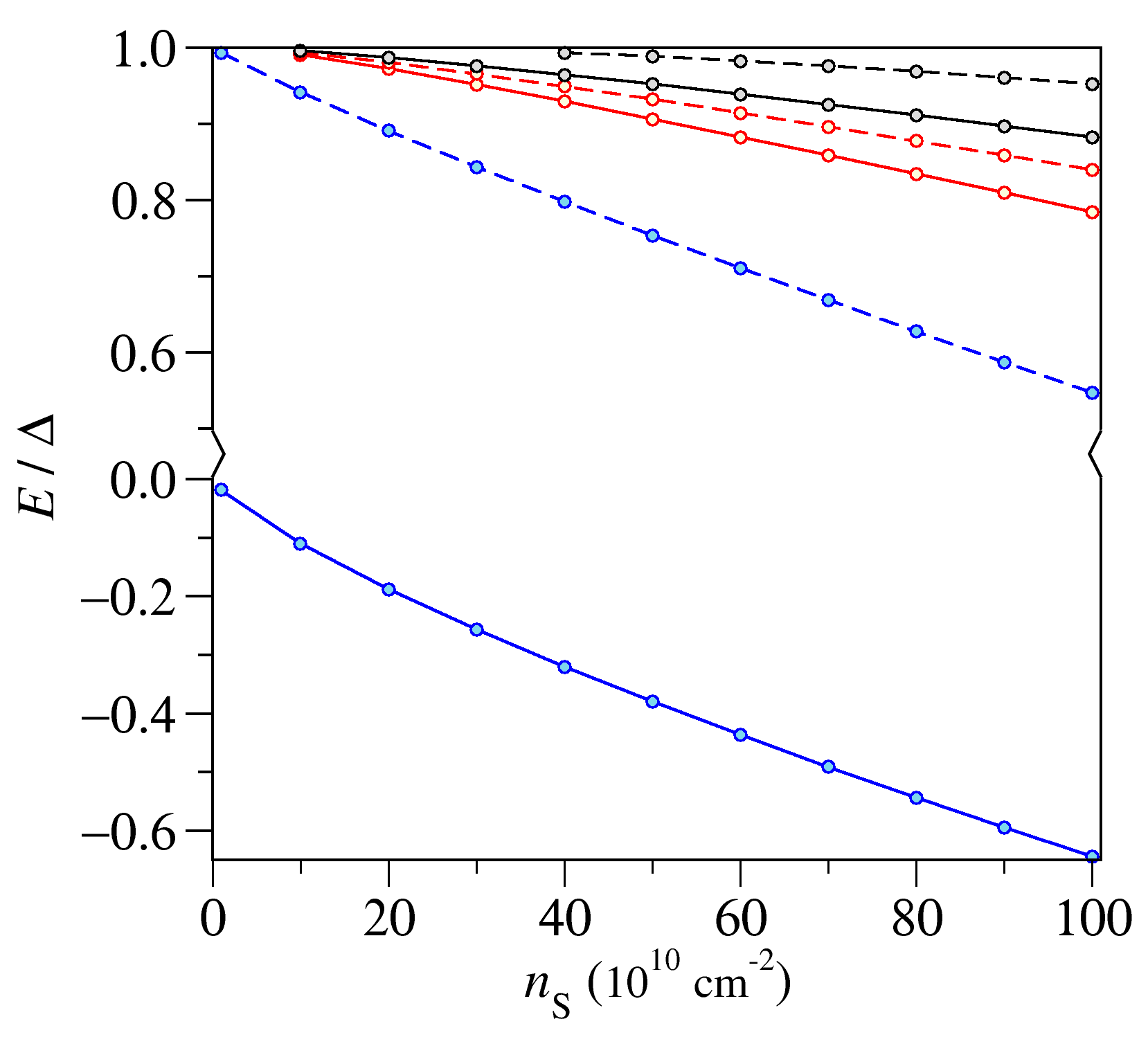}}
\caption{Energy levels as a function of the number of donor impurities per unit area at ${\bm k}=0$. Solid (dashed) lines correspond to band-inverted (normal) systems.}
\label{fig3}
\end{figure}

Our first results are concerned with the evolution of the energy states as a function of doping, $n_{\text{S}}$, for ${\bm k}=0$, as shown in Fig.~\ref{fig3}. As we already discussed in the introduction, the TF well brings states from the continuum into the gap. The TF well localizes the states along the growth direction, although they are extended in the $XY$ plane (they are plane waves). However, when inversion is present, there is already a Dirac state within the energy gap, which prevents continuum states from being hooked by the TF well until the latter is sufficiently strong, that is, until $n_{\text{S}}$ is high enough. As a result, continuum states in the inverted case will enter the gap later than they do in the non-inverted case. In fact, the entering of continuum states of the non-inverted system alternate with those from the inverted one, as displayed in Fig.~\ref{fig3}.

\begin{figure}[ht]
\centering{\includegraphics[width=0.8\columnwidth]{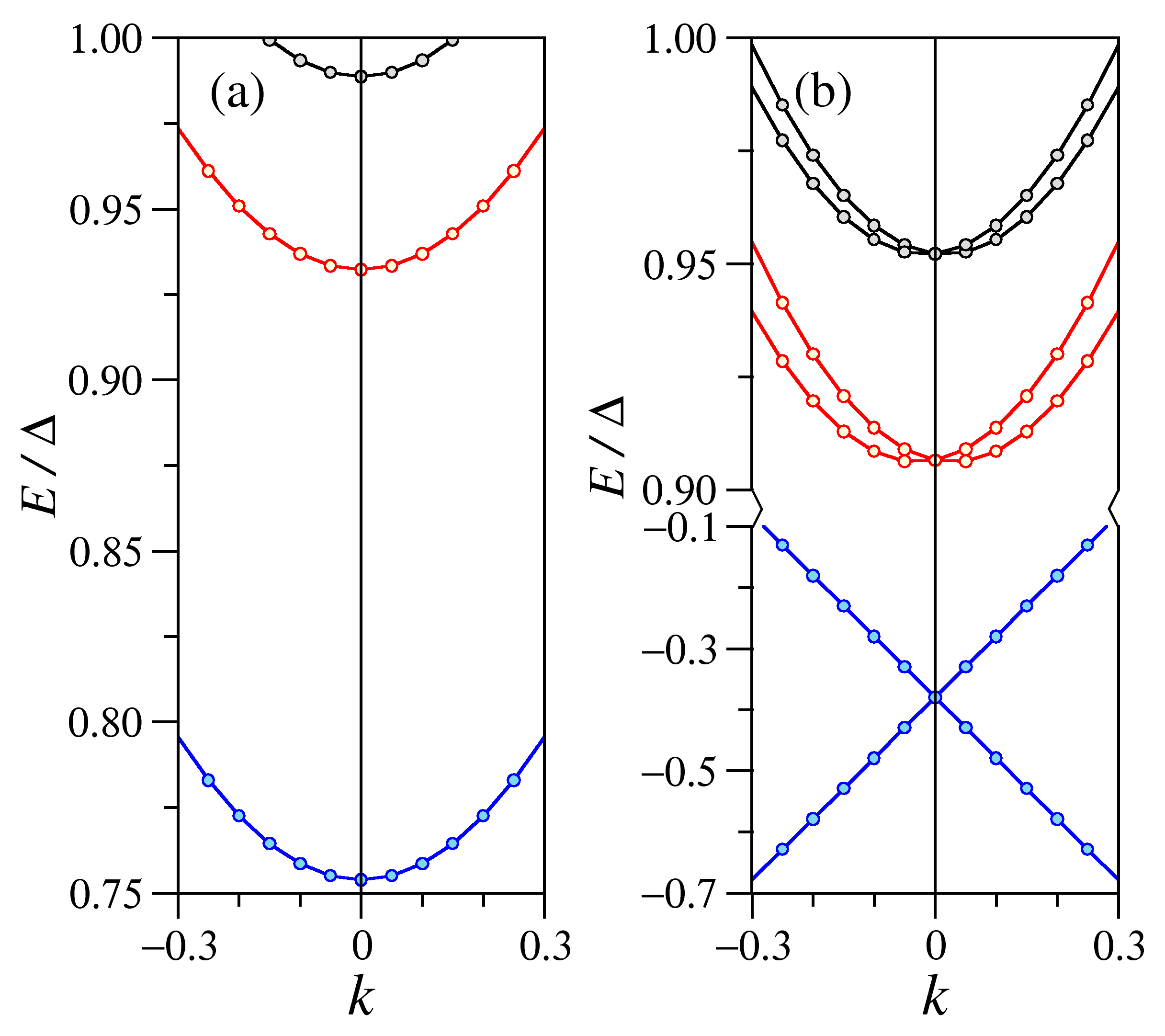}}
\caption{Dispersion relation for (a)~normal and (b)~band-inverted system at $n_\mathrm{S}=\SI{5e11}{\per\centi\metre\tothe{2}}$.}
\label{fig4}
\end{figure}

The next key result comes from studying the dispersion relation, $E(\bm{k})$. Isotropy in the $XY$ plane translates into isotropy in the dispersion relation as well, so we choose an arbitrary direction in ${\bm k}$--space passing through $\bm{k}=0$. This generic direction is denoted by $k$ in the horizontal axis of Fig.~\ref{fig4}. As we can see, massive relativistic dispersion relations are obtained when there is no inversion (see left panel of Fig.~\ref{fig4}). In contrast, when inversion is present, there is a Dirac cone within the gap even in presence of the TF well, an indication of the topological robustness of the cone (see right panel of Fig.~\ref{fig4}). The slope, however, is slightly reduced as compared to the topological boundary without the $\delta$ layer, resembling the result found in biased junctions~\cite{Diaz-Fernandez17a,Diaz-Fernandez17b,Diaz-Fernandez17c,Diaz-Fernandez18}. On the other hand, relativistic massive dispersions entering the gap display a Rashba-like splitting, that is, a horizontal shift of the curves. Although we will not present it here, the splitting can be shown to be a result of mirror symmetry-breaking about $z=0$, that is, due to the presence of an asymmetric boundary, be it topological or not. We have checked numerically that the dispersion curves also split if the energy gaps have the same sign, but their magnitude is different on each side of the boundary.

\begin{figure}[ht]
\centering{\includegraphics[width=0.8\columnwidth]{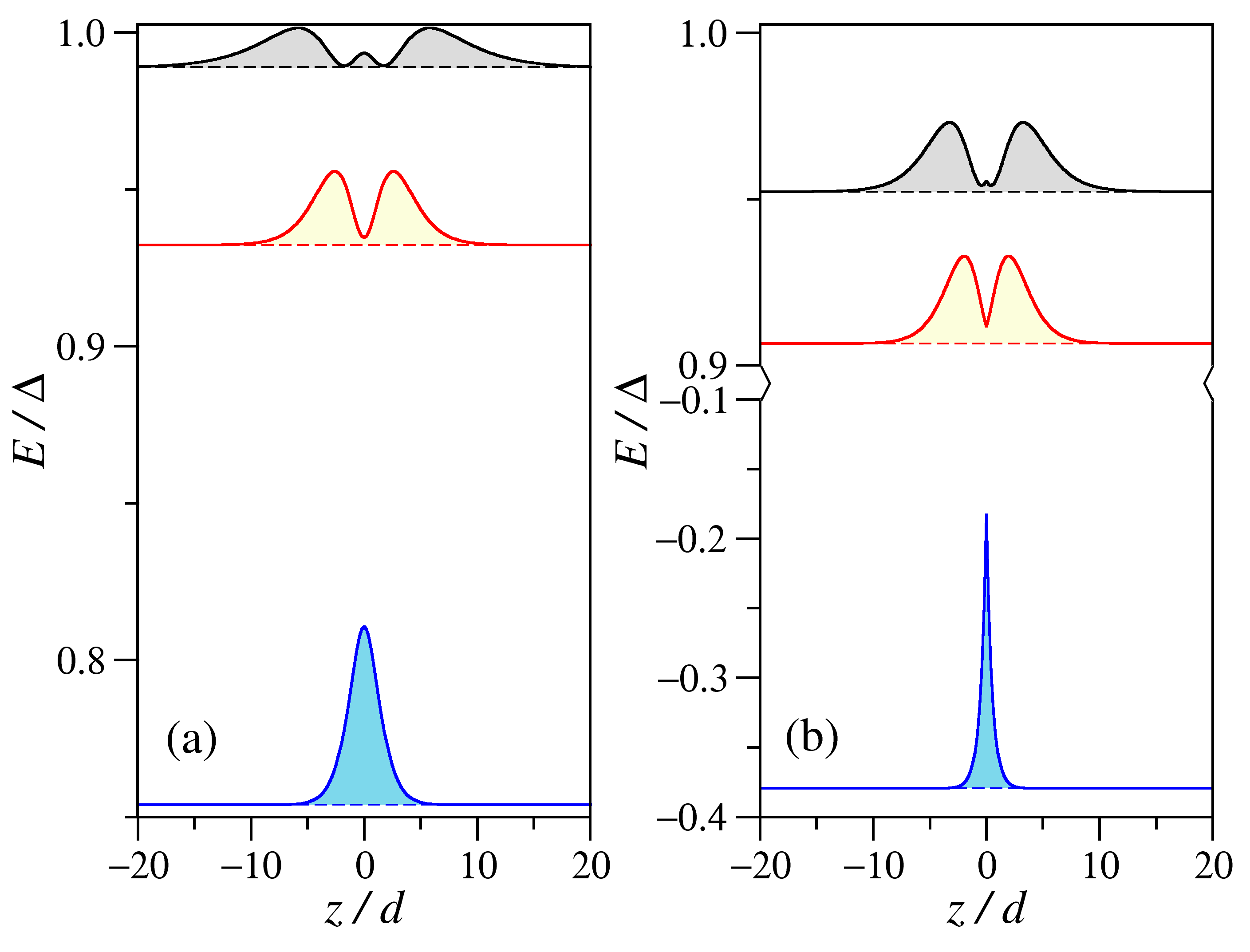}}
\caption{Probability density of the lowest states at ${\bm k}=0$ and $n_\mathrm{S}=\SI{5e11}{\per\centi\metre\tothe{2}}$ for (a)~normal and (b)~band-inverted system. Baselines indicate the energy of the state.}
\label{fig5}
\end{figure}

One can see the localization properties of the envelope function that we discussed at the beginning of this section very easily by looking at the probability density along the growth direction. This is shown in Fig.~\ref{fig5}. If we focus on the more conventional case where there is no inversion (left panel), we can see how the TF well leads to the kind of density profiles that one would expect in an ordinary quantum well, like the bell-shape density profile corresponding to the lowest energy state. More importantly, however, the topological boundary leading to the exponentially localized Dirac state (right panel) dramatically alters the probability density profile of the continuum states entering the gap. For instance, the  topological surface state disallows the first TF well state to be bell-shaped, in contrast to the trivial insulating case. In fact, the hitherto smooth profiles of the TF well now display sharp peaks right at the topological boundary. 

Finally, as we explained in the introduction, optical experiments to detach the response of topological surface states from that of the bulk are said to be difficult to conduct. However, we will now show that a relevant parameter in optical transitions, the oscillator strength, is completely altered when the topological junction is present in contrast to the trivial case. If we denote the initial state by $i$ and the final state by $j$, we can write the oscillator strength as follows~\cite{Peeters93,Davies98}
\begin{equation}
f_{ji} = \frac{2m^{*}\left(E_{j}-E_{i}\right)}{\hbar^2}|\langle j|z|i\rangle|^2 \ ,
\label{eq:13}
\end{equation}
where $m^{*}$ is the effective mass. Using the relation $\Delta=m^{*}v_{\text{F}}^2$~\cite{Korenman87}, the oscillator strength can also be written in terms of the dimensionless variables that we defined earlier in the text as follows $f_{ji} = 2\left(\epsilon_{j}-\epsilon_{i}\right)|\langle j|\xi|i\rangle|^2$. 

\begin{figure}[ht]
\centering{\includegraphics[width=0.8\columnwidth]{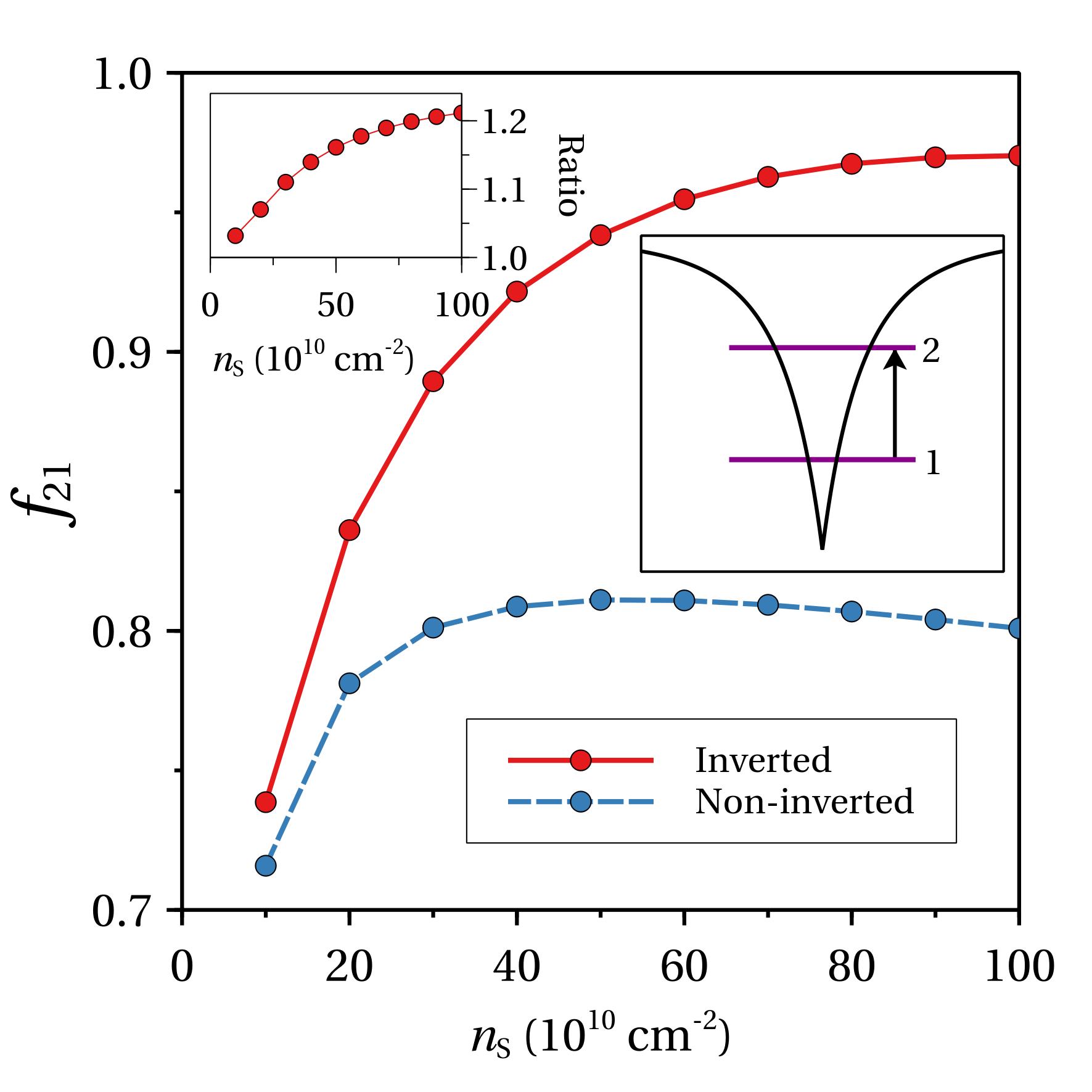}}
\caption{Oscillator strength for optical transitions between the two first states of the TF well for normal (blue line) and inverted (red line) systems as a function of the doping level. The upper left inset displays the ratio of the oscillator strength of the inverted and normal systems.}
\label{fig6}
\end{figure}

In Fig.~\ref{fig6}, we compare the value of the oscillator strength for the transition from the first state of the TF well to the second state at $\bm{k}=0$ as a function of $n_{\text{S}}$, both for the trivial and the topological cases. As it is apparent, the topological boundary has a clear influence on the oscillator strength and, in turn, on the optical response of the system. 
In the trivial system, the oscillator strength reaches its maximum at $n_\mathrm{S}=\SI{5e11}{\per\centi\metre\tothe{2}}$ for the chosen parameters and decreases upon further increase of the doping level. On the contrary, in the topological case, the oscillator strength increases with the doping level in the whole range considered in this work. Most importantly, the oscillator strength is significantly larger in the topological system, up to a $20\%$ as compared to the normal system. Consequently, the intraband optical transitions between the ground and the first excited state of the TF well are enhanced.
Hence, we conclude that optical studies can be carried out in order to efficiently disentangle the response of the surface state from that of the bulk.

\section{Conclusion}

Topological insulators are envisaged to have an ever-increasing number of applications. However, a more complete understanding of the properties of these materials is in order to better exploit these applications. In this work, we seek to unravel some of these fundamental properties. On the one hand, we demonstrate the robustness of the Dirac state against a large perturbation right at the topological boundary, namely, a $\delta$ layer of ionized donor impurities. On the other hand, we show how the linear optical response is markedly reshaped by the presence of the Dirac state. It is our belief that experiments will be able to unfold the optical properties of topological surface states by following the procedure described in this article.

\acknowledgments

The authors thanks P. Rodr\'{i}guez for very enlightening discussions. This research has been supported by MINECO (Grant MAT2016-75955).

\section*{References}

\bibliography{references}

\end{document}